\documentclass[11pt,prd,aps,floats,preprintnumbers]{revtex4}
\usepackage{amsmath,amssymb}
\usepackage{graphicx}
\usepackage{graphics}
\usepackage{epsfig}
\usepackage{latexsym,oldgerm}
\def\beq{\begin{equation}}
\def\eeq{\end{equation}}
\def\bea{\begin{eqnarray}}
\def\eea{\end{eqnarray}}

\newcommand{\comm}[2]{\left[ #1 , #2 \right]}
\newcommand {\nn} {\nonumber}

\newcommand {\bd}{\begin{document}}
\newcommand {\ed}{\end{document}}
\newcommand {\be}{\begin{equation}}
\newcommand {\ee}{\end{equation}}
\newcommand {\ba}{\begin{eqnarray}}
\newcommand {\ea}{\end{eqnarray}}
\newcommand {\bc}{\begin{center}}
\newcommand {\ec}{\end{center}}
\newcommand {\ul}{\underline}
\newcommand {\tc} {\textcolor}

\begin{document}
\small \preprint{SU-4252-897 \vspace{1cm}}
\setlength{\unitlength}{1mm}

\title{The Groenewold - Moyal Plane and its Quantum Physics}

\author{A. P. Balachandran$^{a,b}$}\thanks{bal@phy.syr.edu}\thanks{C\'{a}tedra de
Excelencia}
\author{Pramod Padmanabhan$^{a}$}\thanks{ppadmana@syr.edu}
\affiliation{$^{a}$Department of Physics, Syracuse University,
Syracuse, NY 13244-1130, USA} \affiliation{$^{b}$Departamento de
Matem\'{a}ticas, Universidad Carlos III de Madrid, 28911
Legan\'{e}s, Madrid, Spain}

\begin{abstract}
Quantum theories constructed on the noncommutative spacetime called
the Groenewold-Moyal(GM) plane exhibit many interesting properties
such as causality violation, Lorentz and CPT non-invariance and
twisted statistics. Such violations lead to many striking features
that may be tested experimentally. Thus these theories predict
Pauli-forbidden transitions due to twisted statistics, anisotropies
and acausal effects in the cosmic microwave background radiation in
correlations of observables and Lorentz and CPT violations in
scattering amplitudes. Such features of quantum physics on the GM
plane are surveyed in this review.
\end{abstract}

\maketitle

\section{Introduction}
Physics at the Planck scale can be radically different. This idea is
not new and is attributed to Heisenberg, Pauli and Schr\"{o}dinger.
Later Snyder developed these ideas and published the first paper on
this subject ~\cite{Sny}. It should be noted that Riemann too had
similar visions in the late 19th century when he developed his
geometry. Quantum gravity~\cite{Doplicher} and string theory also
indicate that at very small length scales, spacetime could be
noncommutative.

We here consider a particular model of such a noncommutative algebra
on $\mathbb{R}^{d+1}$ called the Groenewold-Moyal(GM) plane,
$\mathcal{A}_\theta(\mathbb{R}^{d+1})$. This algebra does not admit
the naive action of the Lorentz group. But it does admit a certain
twisted action as we will explain. It results in twisted
statistics~\cite{Apbetal1,Apbetal2} and generalizes the idea of
symmetrized and anti-symmetrized states. These effects can be seen
already at the level of quantum mechanics of multi-particle systems
itself, though they go on to affect quantum field theories on the GM
plane~\cite{cpt}, affecting causality and making them nonlocal
theories.

In this paper we review these developments and also suggest some
future projects. The paper is organized as follows. In section $1$
we discuss the notions of causality and statistics in theories on
commutative spacetimes. In section $2$ we describe the GM plane and
discuss the Moyal star product of functions on this non-commutative
plane. Section $3$ shows how the co-product of spacetime symmetries
gets twisted. The consequence of this twist on the statistics of
multi-particle states is developed in section $4$. In section $5$ we
formulate twisted quantum theories on the GM plane. Sections $6$,
$7$ and $8$ deal with phenomenology on the GM plane illustrating
Pauli forbidden transitions, anisotropies in the CMB spectrum and
violation of CPT. We then present our conclusions in section $9$.

\section{Causality and Statistics}
\label{intro} A crucial ingredient in establishing the connection
between spin and statistics in local quantum field theories is the
requirement of causality \cite{Bogoliubov}, \cite{Haag},
\cite{Weinberg}. This condition of locality, causality is expressed
in such theoretical frameworks by the assumption that the
observables localized at spacelike separated regions commute.

But causality plays a role at much more simple levels. In the theory
of response functions in physical systems, the Kramers-Kronig
relations connect the real and imaginary parts of the response
function by making use of the fact that causality implies
analyticity and vice versa. This is perhaps the simplest context in
non-relativistic physics where causality makes its appearance. We
first recall the derivation of this relation.

A physical system should not respond before the time at which it is
disturbed. Hence if $R(t)$ is the response and disturbance of the
system is zero for time $t<0$,
\begin{equation}
R(t)=0,~~~~t<0.
\end{equation}

Consider its Fourier transform
\begin{equation} \label{R}
\widetilde{R}(\omega)=\int_{-\infty}^\infty dt e^{i\omega t} R(t) =
\int_0^\infty dt e^{i\omega t} R(t)
\end{equation}
For $\textrm{Im}~\omega > 0$, the integral in Eq.(\ref{R}) converges
better because of the extra damping factor $e^{-tIm\omega}$, the
integral in $t$ being from $0$ to $\infty$. Using this fact, one
argues that $\tilde{R}(\omega)$ is holomorphic for $Im\omega > 0$.
This leads to the Kramers-Kronig relations \begin{equation}
{R_1(\omega)=\frac{1}{\pi}\mathcal{P}\int _{-\infty}^{+\infty}
\frac{R_2(\omega')}{\omega'-\omega}d\omega'}\end{equation}
\begin{equation} {R_2(\omega)=-\frac{1}{\pi}\mathcal{P}\int
_{-\infty}^{+\infty}
\frac{R_1(\omega')}{\omega'-\omega}d\omega'}\end{equation} where
$R_1(\omega)$ and $ R_2(\omega)$ denote the real and imaginary parts
of the response function $R(\omega)$ and $\mathcal{P}$ denotes the
Cauchy principal value.

The real and imaginary parts of the response function have physical
interpretations. The imaginary part describes the way the system
dissipates and the real part gives information on scattering
amplitudes.

The Kramers-Kronig relations are the forerunners of dispersion
relations in quantum field theories.

Causal set theory, a discrete approach to quantum gravity, argues
that causality is a ``partial order" and is based on the central
hypothesis that spacetime is a partially ordered or a causal set
\cite{Henson}, \cite{Dowker}, \cite{Sorkin}. In a causal set $C$,
the binary (partial order) condition $\succ$ between its two
elements $x$ and $y$ reads: ``$x \succ y$, if $x$ is to future of
$y$."


There exists a causality condition for the S-matrix called the
Bogoliubov-Shirkov causality condition~\cite{Bogoliubov}. This is
used in axiomatic field theory where the $S$-matrix is a functional
of a function $g$ which is a defined as $g:M\rightarrow [0,1]$,
where $M$ is Minkowskian space. This function measures the amount of
interaction switched on in the action. For $g(x)=0$ there is no
interaction and for $g(x)=1$ the entire interaction is switched on.
For values of $g$ in between the extreme values, the interaction is
only partially switched on. By multiplying the interaction
Lagrangian density $\mathcal{L}(x)$ with $g(x)$ we make the new
action dependent on interaction with intensity $g(x)$. This makes
the scattering matrix a function of $g$. If $x$ and $y$ are two
space-time points and if $x\leq y$ implies that $x$ causally
precedes $y$, then the Bogoliubov-Shirkov causality condition is
\begin{equation}{\frac{\delta}{\delta g(x)}\left(\frac{\delta
S(g)}{\delta g(y)}S^\dagger(g)\right)=0.}\end{equation}

In local quantum field theories, two observable fields $\rho (x)$,
$\eta (y)$ commute if $x$ and $y$ are spacelike separated:
\begin{equation}
\comm{\rho(x)}{\eta(y)}_-=0\end{equation} if
$(x^0-y^0)^2-(\vec{x}-\vec{y})^2<0$, that is, $x\sim y$.

These relations are implemented in quantum field theories by fields.
They need not be observable fields. For scalar fields the above
relation implies that
\begin{equation}
\comm{\varphi(x)}{\chi(y)}_-=0~~~x\sim y,
\end{equation}
and for spinor fields it implies that
\begin{equation}
\comm{\psi^{(1)}_\alpha(x)}{\psi^{(2)}_\beta(y)}_+=0~~~x\sim y,
\end{equation}
where $\pm$ denote anticommutator and commutator respectively. As
these relations also express statistics we see that causality and
statistics are connected.

The above discussion shows that there are different notions of
causality in physics. Their relation is not always clear.

It is interesting to study how the connection between causality and
statistics is affected in quantum field theories that exhibit
features like nonlocality, Lorentz noninvariance etc. A quantum
field theory based on the noncommutativity of spacetime shows these
interesting features. We can model such spacetime noncommutativity
using the algebra of functions called the Groenewold-Moyal (GM)
plane. The GM plane describes noncommutative spacetime where
commutation relations and hence causality and statistics are
deformed.

\section{The Groenewold-Moyal Plane}
The GM plane is the algebra $\mathcal A_\theta$ of smooth functions
on $\mathbb{R}^{d+1}$ with a twisted (star) product. It can be
written as \be f \star g := m_{\theta} (f \otimes g) (x) = m_0
({\cal F}_{\theta} f \otimes g) (x) \ee where $m_0(f \otimes g)(x)
:= f(x) \cdot g(x)$ stands for the usual pointwise multiplication of
the commutative algebra $\mathcal{A}_0$, \be\label{twist} {\cal
F}_{\theta} = \textrm{exp}\Big(\frac{i}{2}\theta^{\mu \nu}
\partial_{\mu} \otimes \partial_{\nu}\Big), \ee is called the
Drinfel'd twist and $\theta ^{\mu \nu} = -\theta ^{\nu \mu}
=\textrm{constant}$. 

The star product implies the commutation relation \be
(\widehat{x}_{\mu} \star \widehat{x}_{\nu} - \widehat{x}_{\nu} \star
\widehat{x}_{\mu}) = \left [ \widehat{x}_\mu, \widehat{x}_\nu \right
]_{\star} = i \theta_{\mu \nu},~~\mu, \nu = 0, 1, \cdots, d, \ee
with $ \widehat{x}_{\mu} = \textrm{coordinate
functions},~~~~\widehat{x}_{\mu}(x) = x_{\mu}.$

We will describe a particular approach to the formulation of quantum
field theories on the GM plane and indicate its  physical
consequences.

It is interesting to see how a noncommutative structure of spacetime
emerges at very small length scales from the arguments based on
Heisenberg's uncertainty principle and Einstein's theory of
classical gravity. Doplicher, Fredenhagen and Roberts
\cite{Doplicher} give the following arguments.

In order to probe physics at the Planck scale $L$, the Compton
wavelength $\hbar/Mc$ of the probe must fulfill $
\frac{\hbar}{Mc}~\leq~L~~~\textrm{or}~~~M~\geq~\frac{\hbar}{Lc}~\simeq~\textrm{Planck
mass}.$ Such high mass in the small volume $L^3$ will strongly
affect gravity and can cause black holes and their horizons to form.
This suggests a fundamental length limiting spatial localization
indicating space-space noncommutativity.

Similar arguments can be made about time-space noncommutativity.
Observation of very short time scales requires very high energies.
If we also try to probe very short length scales at the same time,
then, as in the argument above, they can produce black holes and
black hole horizons will then once more limit spatial resolution
suggesting $\Delta t~\Delta |{\overrightarrow x}|~\geq\
L^2~,~~~L\simeq\textrm{Planck length.}$

The GM plane {\it models} such spacetime uncertainties.

\section{The Twisted Coproduct}
If there is a symmetry group $G$ with elements $g$ and it acts on
single particle Hilbert spaces ${{\cal H}_i}$ by unitary
representations $g \rightarrow U{_i}(g)$, then conventionally it
acts on ${\cal H}_1 \otimes {\cal H}_2$ by the representation
\begin{equation}
g \rightarrow [U_1 \otimes U_2]( g \times g). \label{twoparticle}
\end{equation}

The homomorphism $ \Delta: G \rightarrow G \times G $, ~~  $ g
\rightarrow \Delta(g) := g \times g$ underlying these equations is
said to be a coproduct on $G$.

The action of $G$ on multiparticle states involves more than just
group theory. It involves the coproduct $\Delta$.

The $\star$-multiplication between two functions $f$ and $g$ on the
noncommutative algebra can be expressed in terms of the {\it twist
element}, Eq.(\ref{twist}) \cite{s-matrix}, \cite{Drinfel'd1},
\cite{Chaichian} as
\begin{equation}
f \star g = m_0 \cdot {\cal F}_ \theta (f \otimes g).
\end{equation}

Let $\Lambda$ be an element of the connected component of the
Poincar\'{e} group ${\cal P}_+^\uparrow$. Then for $x \in {\mathbb
R}^N$, we have
\begin{equation}
\Lambda: x \rightarrow \Lambda x \in {\mathbb R}^N.
\end{equation}
It acts on functions on ${\mathbb R}^N$ by pull-back:
\begin{equation}
\Lambda: \alpha \rightarrow \Lambda^* \alpha, \quad (\Lambda^*
\alpha)(x) \alpha[\Lambda^{-1}x].
\end{equation}

The work of Aschieri {\it et al.} \cite{Aschieri} and Chaichian {\it
et al.} \cite{Chaichian} based on Drinfel'd's original work
\cite{Drinfel'd1} shows that ${\cal P}_+^\uparrow$ acts on ${\cal
A}_\theta ({\mathbb R}^N)$ compatibly with $m_\theta$ if its
coproduct is ``twisted'' to $\Delta_\theta$ where
\begin{equation}
\Delta_\theta (\Lambda) = {\cal F}_\theta^{-1} (\Lambda \otimes
\Lambda) {\cal F}_\theta.
\end{equation}

\section{The Twisted Statistics}
The action of the twisted coproduct is not compatible with standard
statistics. Statistics also should be twisted in quantum theory.

A two-particle system for the commutative case ($\theta^{\mu
\nu}=0$) is a function of two sets variables and it lives in
$\mathcal A_0 \otimes \mathcal A_0$. It transforms according to the
usual coproduct $\Delta_0$.

Similarly in the noncommutative case, a state vector lives in
$\mathcal A_\theta \otimes \mathcal A_\theta$ and transforms
according to the twisted coproduct $\Delta_\theta$.

For $\theta^{\mu \nu}=0$, we require that the physical wave
functions describing identical particles are either symmetric
(bosons) or antisymmetric (fermions).

That is, we work with either the symmetrized or antisymmetrized
tensor product
\begin{eqnarray}\label{sa}
\phi \otimes_{S,A} \chi &\equiv& \frac{1}{2}\left(\phi \otimes \chi
\pm\chi \otimes \phi \right).
\end{eqnarray}
The equation (\ref{sa}) is Lorentz invariant.

Similarly in a Lorentz-invariant theory, such relations have to hold
in all frames of reference. But the twisted coproduct action of the
Lorentz group is not compatible with the usual
symmetrization/antisymmetrization. We now discuss this point.

Let $\tau_0$ be the statistics (flip) operator associated with
exchange for $\theta^{\mu \nu}=0$:
\begin{equation}
\tau_0(\phi \otimes \chi) = \chi \otimes \phi.
\end{equation}

For $\theta^{\mu \nu}=0$ , we have the axiom that $\tau_0$ is
superselected. In particular, for Lorentz group action,
$\Delta_{0}(\Lambda) = \Lambda \otimes \Lambda$, must and {\it does
}commute with the statistics operator:
\begin {equation}
\tau_0\ \Delta_{0}(\Lambda)=\Delta_{0}(\Lambda) \tau_0.
\end{equation}
Hence given an element $\phi~\otimes~\chi$ of the tensor product,
the physical Hilbert spaces can be constructed from the elements
\begin{equation}
\left(\frac{1 \pm \tau_0}{2}\right)~(\phi~\otimes~\chi),
\end{equation}
each of them being Lorentz invariant.

Now $\tau_{0} {\cal F}_{\theta} = {\cal F}_{\theta}^{-1} \tau_{0}$
so that $\tau_0\ \Delta_\theta(\Lambda) \neq\ \Delta_\theta(\Lambda)
\tau_0$. This shows that the usual statistics is not compatible with
the twisted coproduct.

But the new statistics operator \cite{statistics-uv-ir}
\begin{equation}
\tau_\theta~\equiv~{\cal F}_\theta^{-1}\tau_0 {\cal F}_\theta, \quad
\tau_\theta^2 = {\bf 1}\otimes {\bf 1}
\end{equation}
does commute with the twisted coproduct $\Delta_{\theta}$:
\begin{equation}
\Delta_\theta (\Lambda) = {\cal F}_\theta^{-1} \Lambda \otimes
\Lambda~{\cal F}_\theta.
\end{equation}

The states constructed according to
\begin{equation}
\phi \otimes_{S_\theta, A_\theta} \chi \equiv \left(\frac{1\,\pm
\tau_\theta}{2}\right)\, (\phi\,\otimes\,\chi),
\end{equation}
form the physical two-particle Hilbert spaces of (generalized)
bosons and fermions respectively and obey twisted statistics.

\section{Twisted Quantum Fields}
A consequence of twisting the co-product of the Poincar\'{e} group
is twisted statistics which in turn reflects on the
commutation(anti-commutation) relations of boson(fermion) fields in
quantum field theory.

A quantum field is an operator-valued distribution acting on a
Hilbert space. We can create a particle localized at a point $x_1$
by acting with a quantum field at $x_1$ on the vacuum. In a similar
way a two-particle state centered at $x_1$ and $x_2$ is created by
acting with the product of two quantum fields at $x_1$ and $x_2$ on
the vacuum.

Consider a free scalar field, $\phi_0$ of mass $m$. We can construct
the following two-particle state: \bea\label{untwisted} \langle
o|\phi_0(x_1)\phi_0(x_2)c^\dagger_\mathbf{q}c^\dagger_\mathbf{p}|0\rangle
& = &(1\pm\tau_0)(e_\mathbf{p}\otimes e_\mathbf{q})(x_1,x_2)
\\ \nn & \equiv & \langle x_1,x_2|p,q\rangle_{S_0,A_0}. \eea
Here $\phi_0(x)$ has the mode expansion
\beq\label{freecf}{\phi_0(x)=\int d\mu(p)(c_{\mathbf{p}}e_p(x) +
d^\dagger_{\mathbf{p}}e_{-p}(x))}\eeq where $e_p(x)=e^{-ip.x} ,
p.x=p_0x_0-\mathbf{p}.\mathbf{x},
d\mu(p)=\frac{1}{(2\pi)^3}\frac{d^3p}{2p_0},
p_0=\sqrt{\mathbf{p}^2+m^2}$. The creation and annihilation
operators satisfy the standard commutation(anti-commutation)
relations, \beq{c_{\mathbf{p}}c_{\mathbf{q}}^\dagger \pm
c_{\mathbf{q}}^\dagger c_{\mathbf{p}} = 2p_0
\delta^3({\mathbf{p}}-{\mathbf{q}}),~~d_{\mathbf{p}}d_{\mathbf{q}}^\dagger
\pm d_{\mathbf{q}}^\dagger d_{\mathbf{p}} = 2p_0
\delta^3({\mathbf{p}}-{\mathbf{q}}).}\eeq

The two-particle states in non-commutative quantum field theory
should obey twisted statistics. Using Eq.(\ref{freecf}) as the
guiding principle we can construct the twisted scalar quantum field
$\phi_{\theta}(x)$ as \beq{\phi_{\theta}(x)=\int
d\mu(p)(a_{\mathbf{p}}e_p(x) + b^\dagger_{\mathbf{p}}e_{-p}(x))}\eeq
The creation and annihilation operators defined in the above mode
expansion are twisted as we shall see below.

Using the above twisted field we can construct two-particle states
as in Eq.(\ref{untwisted}):  \bea \langle
o|\phi_{\theta}(x_1)\phi_{\theta}(x_2)a^\dagger_\mathbf{q}a^\dagger_\mathbf{p}|0\rangle
& = &(1\pm\tau_{\theta})(e_\mathbf{p}\otimes e_\mathbf{q})(x_1,x_2)
\\ \nn & \equiv & \langle x_1,x_2|p,q\rangle_{S_{\theta},A_{\theta}}. \eea
From these relations we get the relation ~(see \cite{cpt} for more
details) \beq{a_{\mathbf{p}}^\dagger a_{\mathbf{q}}^\dagger=\pm
e^{ip_{\mu}\theta^{\mu\nu}q_{\nu}}a_{\mathbf{q}}^\dagger
a_{\mathbf{p}}^\dagger, ~~a_{\mathbf{p}} a_{\mathbf{q}}=\pm
e^{ip_{\mu}\theta^{\mu\nu}q_{\nu}}a_{\mathbf{q}} a_{\mathbf{p}}}\eeq

It is possible to write the twisted creation and annihilation
operators $a_{\mathbf{p}}^\dagger, a_{\mathbf{p}}$ in terms of the
untwisted operators in Eq.(\ref{freecf}). The transformation
connecting the twisted and untwisted creation and annihilation
operators is called the ``dressing transformation" \cite{dressing1,
dressing2} and is given by
\beq{a_{\mathbf{p}}=c_{\mathbf{p}}e^{-\frac{i}{2}p_{\mu}\theta^{\mu\nu}P_{\nu}},~~b_{\mathbf{p}}=d_{\mathbf{p}}e^{-\frac{i}{2}p_{\mu}\theta^{\mu\nu}P_{\nu}}.}\eeq
Here $P_{\mu}$ is the four-momentum operator given by $\int
\frac{d^3p}{2p_0}(c_{\mathbf{p}}c_{\mathbf{p}}^\dagger+d_{\mathbf{p}}d_{\mathbf{p}}^\dagger)p_{\mu}$.
Note here that we have written the four-momentum operator in terms
of the untwisted creation and annihilation operators. The expression
for the four-momentum in terms of the twisted ones are just got by
replacing the untwisted by the twisted ones as
$p_{\mu}\theta^{\mu\nu}P_{\nu}$ commutes with $c_{\mathbf{p}},
c_{\mathbf{p}}^\dagger$.

We can write the twisted quantum field in terms of the untwisted one
with the help of the dressing transformation as
\beq{\phi_{\theta}(x)=\phi_0(x)e^{\frac{1}{2}\overleftarrow{\partial_{\mu}}\theta^{\mu\nu}P_{\nu}}}\eeq

\section{The Pauli Principle}
In \cite{Chakraborty} the statistical potential
$V_{\textrm{\tiny{STAT}}}$ between two identical fermions at inverse
temperature $\beta$ has been computed:
\begin{equation}
{\textrm{exp}}\Big(-\beta V_{\textrm{\tiny{STAT}}}
({\overrightarrow{\bf x}}_{1}, {\overrightarrow{\bf x}}_{2})\Big) =
\langle {\overrightarrow{\bf x}}_{1}, {\overrightarrow{\bf
x}}_{2}|e^{-\beta H}|{\overrightarrow{\bf
x}}_{1},{\overrightarrow{\bf x}}_{2}\rangle,~~H =
\frac{1}{2m}({\overrightarrow{\bf p}}_{1}^{2}+{\overrightarrow{\bf
p}}_{2}^{2}).\nonumber
\end{equation}
Here $|{\overrightarrow{\bf x}}_{1},{\overrightarrow{\bf
x}}_{2}\rangle$ has twisted antisymmetry:
$\tau_{\theta}|{\overrightarrow{\bf x}}_{1},{\overrightarrow{\bf
x}}_{2}\rangle = - |{\overrightarrow{\bf
x}}_{1},{\overrightarrow{\bf x}}_{2}\rangle$. It is explicitly shown
not to have an infinitely repulsive core, establishing the violation
of Pauli principle, as has been earlier suggested ~\cite{ApbGmSvAp}.

This result has phenomenological consequences such as Pauli
forbidden transitions (on which there are stringent limits). We
indicate them below:

Non-Paulian atoms or nuclei are those whose orbitals can be filled
with extra electrons or nuclei violating the Pauli principle. As an
example, non-Paulian carbon has as its atomic configuration, $1s^3
2s^2 2p^1$. The presence of just a single electron in the outermost
shell of non-Paulian carbon makes it behave chemically like boron,
whose atomic configuration is $1s^2 2s^2 2p^1$. Thus by searching
for non-Paulian carbon atoms in samples of boron, we can get the
concentration of the former in the latter. Bounds on these values
were found by the NEMO experiments~\cite{Barabash}.

In the Borexino \cite{Back:2004hd} and SuperKamiokande
\cite{Suzuki:1993zp} experiments, the forbidden transitions from
$O^{16}$($C^{12}$) to $\tilde{O}^{16}$($\tilde{C}^{12}$) where the
tilde nuclei have an extra nucleon in the filled $1S_{1/2}$ level
are found to have lifetimes greater than $10^{27}$ years. There are
also experiments on forbidden transitions to filled K-shells of
crystals done in Maryland which give branching ratios less than
$10^{-25}$ for such transitions. The consequences of these results
for noncommutative models are yet to be studied.

\section{Cosmic Microwave Background (CMB)}
The COBE satellite, in 1992, detected anisotropies in the CMB
radiation, which led to the conclusion that the early universe was
not smooth: there were small perturbations in the photon-baryon
fluid.

The perturbations could be due to the quantum fluctuations in the
inflaton (the scalar field driving inflation). These fluctuations
act as seeds for the primordial perturbations over the smooth
universe. Thus according to these ideas, the early universe had
inhomogeneities and we observe them today in the distribution of
large scale structure and anisotropies in the CMB radiation.

The temperature field in the sky can be expanded in spherical
harmonics: \be {\Delta T(\hat{n}) \over T} = \sum_{l m} a_{l m} Y_{l
m}(\hat{n}). \ee

The $a_{lm}$ can be written in terms of perturbations of the
Newtonian potential $\Phi$ \be a_{lm} =
4\pi(-i)^l\int\frac{d^3k}{(2\pi)^3}~\Phi(k)\Delta^T_l(k)Y^\ast_{lm}(\hat{k}),
\ee where $\Delta^T_l(k)$ are called the transfer functions.

In a noncommutative spacetime, the quantum corrections to the
inflaton fluctuations are modified. The angular correlation
functions $\langle a_{lm} a^{*}_{l'm'}\rangle_{_\theta}$ acquire
non-diagonal elements in the magnetic quantum number indicating
rotational symmetry breaking in the universe.

It is also $\theta$ dependent indicating a preferred direction and
are not invariant under rotations. The correlations of $a_{lm}$ are
not Gaussian either~\cite{cmbpaper}.

On fitting data \cite{cmbpaper2}, one finds an upper bound for the
length and energy scales associated with spacetime noncommutativity,
\be \sqrt{\theta} \lesssim 10^{-17} \textrm{cm},~~ E \gtrsim 10^{3}
\textrm{GeV}. \ee

\section{Causality, Lorentz invariance and CPT}

{\it i)} \textit{Causality and Lorentz Invariance}: The $S$-matrix
of quantum theories constructed on the GM plane is not Lorentz
invariant. The underlying reason is loss of causality. Causality is
known to be generally required for the Lorentz invariance of the
$S$-matrix~\cite{Bogoliubov, Weinberg}.

Thus let ${\cal H}_I$ be the interaction Hamiltonian density in the
interaction representation of the quantum theory. The interaction
representation $S$-matrix is
\begin{equation}
S = T \exp \left( -i \int d^4 x {\cal H}_I(x) \right).
\end{equation}
Bogoliubov and Shirkov \cite{Bogoliubov} and then Weinberg
\cite{Weinberg} long ago deduced from causality (locality) and
relativistic invariance that ${\cal H}_I$ must be a local field:$
[{\cal H}_I(x), {\cal H}_I(y)] = 0,~~~x \sim y$. But noncommutative
theories are nonlocal and violate this condition: this is the
essential reason for Lorentz noninvariace.

The effect of Lorentz noninvariace on scattering amplitudes is
striking. They depend on the total incident momentum
$\overrightarrow{P}_{\textrm{inc}}$ through the term
$\theta^{0i}P^{\textrm{inc}}_{i}$.

The effects of $\theta^{\mu \nu}$ disappear in the center-of-mass
system, or more generally if $\theta^{0i}P^{\textrm{inc}}_{i} = 0$.
But otherwise there is dependence on $\theta_{0i}$.

{\it ii)} \textit{CPT}: The noncommutative $S$-matrix transforms
under CPT in the following way \cite{cpt}: \ba {S}^{^{M,
G}}_{\theta} &=& \text{T}~\exp~\Big[-i\int d^{4}x~{\cal H}^{^{M,
G}}_{I
0}(x)~e^{{1 \over 2}\overleftarrow{\partial}\wedge P}\Big] \nn \\
&&\rightarrow \text{T}~\exp~\Big[i\int d^{4}x~{\cal H}^{^{M, G}}_{I
0}(x)~e^{-{1 \over 2}\overleftarrow{\partial}\wedge P}\Big] =
({S}^{^{M, G}}_{-\theta})^{-1}. \ea Here ${\cal H}^{^{M, G}}_{I 0}$
is the matter-gauge interaction Hamiltonian density for $\theta^{\mu
\nu}=0$. After performing the spatial integration, we can reduce
$e^{\frac{1}{2}\overleftarrow{\partial}\wedge P}$ in the $S$-matrix
to $e^{\frac{1}{2}\overleftarrow{\partial}_{0} \theta^{0i} P_{i}}$.
Thus the effect of P and CPT is to reverse the sign of
$\theta^{0i}$: P or CPT :~~$\theta^{0i} \rightarrow -\theta^{0i}$.
The $\theta^{0i}$ contributes to P, and more strikingly, to CPT
violation.

The particle-antiparticle life times can differ to order
$\theta^{0i}$ because of CPT violation: \be
\tau_{\rm{\tiny{particle}}} - \tau_{\rm{\tiny{antiparticle}}} \cong
\theta^{0i} P^{\textrm{inc}}_{i}. \ee It can give rise to
interesting effects such as mass difference in $K^0 - \bar{K}^0$
system and the difference in $(g-2)$ of $\mu^{\pm}$. (See
\cite{Joseph:2008fz} for bounds on $\theta$ estimated from these
effects.)

{\it Remark}: The decay $Z^{0} \longrightarrow 2 \gamma $ is
forbidden even with noncommutativity in the approach of Aschieri
{\it et. al.} More generally, a massive particle of spin $j$ does
not decay into two massless particles of same helicity if $j$ is
odd. This result is the extension of the Landau-Yang
theorem~\cite{Landau}. The search for this event is thus of
fundamental significance.

\section{Conclusions}
Spacetime noncommutativity deforms statistics and so generically
violates causality in noncommutative quantum theories. Such effects
lead to many interesting features such as $i)$ modification of the
Pauli principle causing forbidden atomic transitions, $ii)$
correlations of observables in spacelike regions giving rise to
anisotropies in the CMB radiation, and $iii)$ Lorentz and CPT
violations in scattering amplitudes.

In this review, we have discussed certain specific observable
predictions of these effects and derived bounds on the
noncommutativity parameter $\theta_{\mu\nu}$. However more
theoretical and experimental work is needed to obtain strong and
reliable bounds on $|\theta_{\mu\nu}|$. \\

\textbf{Acknowledgements}\\

APB thanks Alberto Ibort and the Universidad Carlos III de Madrid
for their kind hospitality and support. This work was partially
supported by the US Department of Energy under grant number
DE-FG02-85ER40231. This work is based on the talk given by APB at
the XXV Max Born Symposium on Planck Scale Physics, Wroclaw, $29$
June-$3$ July $2009$.

\end{document}